\documentclass[conference]{IEEEtran}
\IEEEoverridecommandlockouts
\usepackage{graphicx}
\usepackage{amsmath}
\usepackage{fancyhdr}
\usepackage{hyperref}
\usepackage{authblk}
\usepackage{xcolor}

\ifCLASSINFOpdf
\else
\fi

\begin{document}
%
\title{Characterizing and Subsetting Big Data Workloads}

\author[1,2]{Zhen Jia}
\author[1*]{Jianfeng Zhan \thanks{* The corresponding author is Jianfeng Zhan.}}
\author[1]{Lei Wang}
\author[1]{Rui Han}
\author[3]{Sally A. McKee}
\author[1]{Qiang Yang}
\author[1]{Chunjie Luo}
\author[1]{Jingwei Li}
\affil[1]{State Key Laboratory Computer Architecture, Institute of Computing Technology, Chinese Academy of Sciences}
\affil[2]{University of Chinese Academy of Sciences, China}
\affil[3]{Chalmers University of Technology, Sweden}
\affil[  ]{Email: \{jiazhen, zhanjianfeng, wanglei\_2011, hanrui\}@ict.ac.cn, mckee@chalmers.se,\authorcr \{yangqiang,luochunjie,lijingwei\}@ict.ac.cn}

\maketitle

\begin{abstract}
Big data benchmark suites must include a diversity of data and workloads
to be useful in fairly evaluating big data systems and architectures.
However, using truly comprehensive benchmarks poses great challenges
for the architecture community. First, we need to thoroughly understand the
behaviors of a variety of workloads. Second, our usual simulation-based
research methods become prohibitively expensive for big data. As big data is an
emerging field, more and more software stacks are being proposed to
facilitate the development of big data applications, which aggravates these challenges.

In this paper, we first use
Principle Component Analysis (PCA) to identify the most important characteristics
from 45  metrics to characterize big data workloads from BigDataBench,
a comprehensive big data benchmark suite. Second, we apply a clustering technique
to the principle components obtained from the PCA to investigate the similarity
among big data workloads, and we verify the importance of including different
software stacks for big data benchmarking. Third, we select
seven representative big data workloads by removing redundant ones and
release the BigDataBench simulation version, which is publicly available from \url{http://prof.ict.ac.cn/BigDataBench/simulatorversion/}.

\end{abstract}

\IEEEpeerreviewmaketitle
\section{Introduction}
Today, huge amounts of data are being collected in many areas,
creating new opportunities to understand phenomena in meteorology,
health, finance, and many other sectors. Every two days we create as
much information as we did from the dawn of civilization up until 2003,
and the pace is increasing~\cite{Eric}. Reports from IDC forecast that
from 2005 to 2020, the digital universe will grow by a factor of 300,
from 130 exabytes to 40,000 exabytes (i.e., 40 trillion gigabytes)~\cite{IDC2020}.
Turning such big data into valuable information requires the support
of big data systems, and the design of such systems is a
growing research topic in both academia and industry. In this context,
the pressure to develop innovative theories and technologies to
improve the performance, energy-efficiency, and cost-effectiveness of big systems rises.

The cornerstone of such research are big data benchmarks that facilitate accurate evaluation
of big data systems and a better understanding of the behaviors of big data
workloads (applications). Fair evaluation requires diversity in both the data sets
and the workloads used in benchmarking big data systems and architectures.
Generating such comprehensive and representative benchmarks raises
two great challenges for the architecture community. First, the plethora of
microarchitectural metrics that can be tracked creates a potentially huge
amount of characterization data that can be hard to analyze.
Second, the simulation-based approaches that are widely used in
architecture research  are very time-consuming: simulating complete workloads
amounts to an impossible mission.
How to address these challenges remains an open question.

More and more software stacks
are being proposed to facilitate the development of big data applications.
Previous work~\cite{zaharia2012resilient}~\cite{AMP} has shown that software stacks
can cause big data workloads to have different user-observed performance.
For instance, compared to Hadoop, Spark improves runtime performance by factors
of up to 100. Different software
stacks should thus be included in the benchmarks,
but that aggravates the above challenges by
multiplying the number of workloads.

To date, the major efforts on big data benchmarking
either propose a comprehensive but a large amount of workloads 
or only select a few workloads according to so-called popularity~\cite{ferdman2011clearing}. And hence, achieving benchmark comprehensiveness while enabling
efficient experimentation is particularly
acute in big data systems research.
Although many workload subsetting approaches
have been
proposed~\cite{eeckhout2003quantifying,yi2006evaluating,eeckhout2002workload,phiansalkar2007analysisab},
these are designed for workloads
having significantly different characteristics from big data applications.

In this paper, we propose a general approach to
1) identify the most important metrics for characterizing big data
workloads and 2) cull redundant workloads.
To demonstrate the
effectiveness of the approach, we select two prevalent big data processing
software stacks (Hadoop~\cite{white2009hadoop} and Spark~\cite{zaharia2012resilient}), single out 32 workloads from
BigDataBench~\cite{wang2014bigdatabench}
to run on these stacks, and evaluate them via 45 typical microarchitectural
metrics that represent basic characteristics of modern processors.
After applying Principle Component Analysis (PCA) and hierarchical clustering, we find the following:

\textbf{Software stacks have significant impact on the microarchitectural
behaviors of big data workloads.} This finding is based on the five observations
in Section~\ref{macro}. We observe that workloads implementing different
algorithms running on the same software stack are more likely to have similar behaviors
than same-algorithm workloads running on
different software stacks (in the first clustering iteration,
80\% of clusters consist of workloads that are based on
the same software stack). This indicates that (in our experimental setting,
at least) the impact of software stacks is greater than
that of algorithms.
Hence, it is reasonable to say that different software stacks must be included
in performing representative and credible evaluations of big data systems. 
Compared to traditional software stacks, big data
software stacks usually have more complex structures.
Such designs enable programmers to write
less code to achieve their intended goals.   The upshot is that the ratio of system
software and middleware instructions executed compared to
user applications instructions tends to be large, which makes their impact
on system behavior large, as well.

\textbf{We can identify the most important microarchitectural-level metrics for studying
the impact of different software stacks.} We apply PCA
to 45 microarchitectural metrics, finding that the L3 cache miss rate, instruction
fetch stalls, data TLB behaviors, and snoop responses are the most
important metrics in differentiating Hadoop-based and Spark-based workloads.

\textbf{We can successfully subset big data workloads.}
Based on our derived principle components,
we employ another clustering technique to select seven representative
workloads (out of the original 32) by retaining those exhibiting unique
behaviors.
We also use a statistical method (Bayesian
Information Criterion) to ensure that the
clustering is a ``good fit'' to our workloads.
We deploy the representative workloads on a full-system simulator and
release the BigDataBench simulator version, which is publicly available from \url{http://prof.ict.ac.cn/BigDataBench/simulatorversion/}.
\section{Background} \label{back}
Measuring big data systems and architectures quantitatively
is the foundation of innovative big data research, and a
good benchmark suite can make this task much more efficient.
When constructing such benchmarks,
making their behavior representative of real workloads
is important to their success~\cite{gray1992benchmark,bienia2008parsec}.
We choose BigDataBench~\cite{wang2014bigdatabench}, a recent
benchmark suite, for our workload
characterization for the following reasons:

\begin{enumerate}
\item
It covers representative application domains. BigDataBench includes
microbenchmarks and applications from search engines, social networks,
and e-commerce --- i.e., the most important internet services
from typical big data application domains.

\item
It covers representative workloads. As in the widely used TPC-C benchmark\cite{chen2014tpc},
units of computation that appear frequently in the
benchmarked application domain are used in benchmark construction.
\item It covers diverse software stacks \cite{wang2010transformer}. In BigDataBench,
the offline analytics workloads use Hadoop and Spark,
and the interactive analytics and OLAP workloads use Shark, Hive, and etc.

\item It considers data volume and veracity.
BigDataBench develops Big Data Generator Suite (BDGS), an open source
tool to generate synthetic big data based on six raw data sets~\cite{ming2014bdgs}.
\end{enumerate}

\section{Methodology}
We use the following methodology to
analyze the software stack's impact on big data workloads
from a microarchitectural perspective. First we select representative
workloads in big data fields.
Then we identify a set of microarchitectural metrics that
can directly or indirectly reflect
program behavior.
Finally, we use standard statistical methods in our
characterizations.

\subsection{Workload Selection}\label{workloadsSelect}

We select BigDataBench workloads that include
structured, semi-structured, and unstructured data.
To minimize the impact of
non-workload factors,
we ensure the following properties in our comparisons.

\textbf{Identical Algorithms.} All applications we
select have two implementations
of the same algorithm --- a Hadoop-based implementation
and a Spark-based implementation.
Table~\ref{workloads} lists the algorithms we use.
Recall that the offline analytics workloads
contain algorithms implemented directly on Hadoop or Spark,
and the interactive analytics workloads involve SQL-like operations
on Hive~\cite{thusoo2009hive} or Shark~\cite{engle2012shark}.
Hive operations are interpreted in Hadoop jobs, and
Shark operations are interpreted in Spark jobs.
Table~\ref{workloads} lists the workloads we use in this paper.

\textbf{Identical Data Sets.} We use the same data set to drive
both the Hadoop-based and Spark-based workloads, i.e.,
both data formats and data sizes are identical.
Table~\ref{workloads} lists these data sets and formats.
The input data size for each workload
is determined by the Spark-based implementation,
since Spark is an in-memory computing engine.
Both the input and intermediate data should fit in
memory to attain reasonable experiment time.
For each Spark-based workload, we test several input data sets,
ranging from several gigabytes to more than 100 gigabytes. 

\textbf{Same Infrastructure.} Both workload implementations
run on the same cluster under the same software
environment (i.e., with the same numbers of threads, the
same OS version, and the same JVM version).
Section~\ref{exeevn} gives configuration details.

\begin{table}[hbtp]
\caption{Representative Data Analysis Workloads}\label{workloads}
{
\scriptsize
\centering
\begin{tabular}{|p{1cm}|p{1.5cm}|p{1.2cm}|p{1.5cm}|p{2.0cm}|c|} \hline
 Category &Workload  &  Problem Size &Data Type &  Software Stack \\ \hline
 &Sort     & 80 GB   & unstructured  sequence file &\\ \cline{2-4}
&WordCount&  98 GB &unstructured text  &\\ \cline{2-4}
 Offline Analytics  &Grep  & 98 GB  & unstructured text   & Hadoop \& Spark \\ \cline{2-4}
&Naive Bayes& 84 GB & semi-structured text &  \\ \cline{2-4}
&K-means  & 44 GB & unstructured text  &\\ \cline{2-4}
&PageRank  & $2^{24}$ vertices  & unstructured graph  &\\ \hline
&Projection& 420 million records  &     &\\ \cline{2-3}
&Filter  &420 million records &  &\\ \cline{2-3}
&Order By  & 420 million records &    &\\ \cline{2-3}
Interactive Analytics &Cross product  & 100 million records &    &Hive \& Shark\\ \cline{2-3}
&Union  & 420 million records& structured table  &\\ \cline{2-3}
&Difference  &100 million records & (e-commerce transaction data set) &\\ \cline{2-3}
&Aggregation  &420 million records & &\\  \cline{2-3}
&JoinQuery & 100 million records&  &\\ \cline{2-3}
&AggQuery & 420 million records & &\\ \cline{2-3}
&SelectQuery & 420 million records & &\\\hline
\end{tabular}
}
\end{table}

\subsection{Microarchitectural Metric Selection} \label{eMethodology}
Good workload characterization incorporates a variety of metrics
to better gain a comprehensive understanding of the target programs~\cite{eeckhout2002workload}.
To analyze microarchitectural behaviors, we choose a broad set of metrics of different types
that cover all major characters.
We particularly focus on factors that may affect data movement or calculation.
For example, a cache miss may delay data movement, and a branch misprediction
flushes the pipeline.
Table~\ref{metric} summarizes them, and we categorize them below.

\textbf{Instruction Mix.} The instruction mix can reflect a workload's logic
and affect performance.
Here we consider both the instruction type and the execution mode (i.e., user mode
running in ring three and kernel mode running in ring zero).

\textbf{Cache Behavior.}
The processor in our experiments has private L1 and L2 caches per core, and all
cores share an L3. The L1 cache is split for instructions and data.
The L2 and L3 are unified.
We track the cache misses per kilo instructions and
cache hits per kilo instructions except L1 data cache, noting that for the L1D
miss penalties may be hidden by out-of-order cores~\cite{karkhanis2004first}.

\textbf{Translation Look-aside Buffer (TLB) Behavior.}
Our processor has a two-level TLB. The first level has separate
instruction and data TLBs. The second level is shared.
We collect statistics at both levels.

\textbf{Branch Execution.}
We consider the miss prediction ratio and the ratio of
branch instructions executed to those retired. These reflect
how many branch instructions are predicted wrong and how many are flushed.

\textbf{Pipeline Behavior.}
Stalls can happen in any part of the pipeline, but
superscalar out-of-order  processors prevent us from precisely
breaking down the execution time~\cite{ferdman2011clearing,keeton1998performance,eyerman2006performance}.
Retirement-centric analysis also has difficulty
accounting for how the CPU cycles are used because the
pipeline continues executing instructions even when
retirement is blocked~\cite{levinthal18027cycle}.
Here we focus on counting
cycles stalled due to resource conflicts, e.g.,
reorder buffer full stalls that prevent new instructions from
entering the pipeline.

\textbf{Offcore Requests and Snoop Responses.} Offcore requests
tell us about individual core requests to the LLC (Last Level Cache).
Requests can be classified into data requests, code requests,
data write-back requests, and request for ownership (RFO) requests.
Snoop responses give us information on the workings of the
cache coherence protocol.

\textbf{Parallelism.} We consider Instruction Level Parallelism (ILP)
and Memory Level Parallelism (MLP). ILP reflects how many instructions
can be executed in one cycle (i.e., the IPC),
and MLP reflects how many outstanding cache requests are
being processed concurrently.

\textbf{Operation Intensity.} The ratio of computation to memory accesses
reflects a workload's computation pattern. For instance, most big data
workloads have a low ratio of floating point operations to memory accesses,
whereas HPC workloads generally have high floating point operations to memory  accesses
ratios~\cite{wang2014bigdatabench}.

\begin{table}
\caption{microarchitecture Level Metrics.}\label{metric}
\center
\begin{tabular}{|p{1cm}|c|c|p{3.5cm}|}
  \hline
  Category & No. & Metric Name & Description \\ \hline
  Instruction  & 1 & LOAD& load operations' percentage \\ \cline{2-4}
  Mix & 2 & STORE & store operations' percentage \\ \cline{2-4}
   & 3 & BRANCH & branch operations' percentage \\ \cline{2-4}
   & 4 & INTEGER & integer operations' percentage\\ \cline{2-4}
   & 5 & FP & X87 floating point operations' percentage\\ \cline{2-4}
   & 6 & SSE FP & SSE floating point operations' percentage \\ \cline{2-4}
   & 7 & KERNEL MODE & the ratio of instruction running on kernel mode \\ \cline{2-4}
   & 8 & USER MODE & the ratio of instruction running on user mode \\ \cline{2-4}
   & 9 & UOPS TO INS & the ratio of micro operation to instruction \\ \hline
Cache Behavior & 10 & L1I MISS & L1 instruction cache misses per K instructions \\ \cline{2-4}
 & 11 & L1I HIT & L1 instruction cache hits per K instructions \\ \cline{2-4}
      & 12 & L2 MISS & L2 cache misses per K instructions \\ \cline{2-4}
      & 13 & L2 HIT & L2 cache hits per K instructions \\ \cline{2-4}
      & 14 & L3 MISS & L3 cache misses per K instructions \\ \cline{2-4}
      & 15 & L3 HIT & L3 cache hits per K instructions \\ \cline{2-4}
      & 16 & LOAD HIT LFB  & loads miss the L1D and hit line fill buffer per K instructions \\ \cline{2-4}
      & 17 & LOAD HIT L2 & loads hit L2 cache per K instructions\\ \cline{2-4}
      & 18 & LOAD HIT SIBE & loads hit sibling core's L2 cache per K instructions\\ \cline{2-4}
 & 19 & LOAD HIT L3 & loads hit unshared lines in L3 cache per K instructions\\ \cline{2-4}
 & 20 & LOAD LLC MISS & loads miss the L3 cache per K instructions\\ \hline
 TLB Behavior & 21 & ITLB MISS & misses in all levels of the instruction TLB  per K instructions\\ \cline{2-4}
& 22 & ITLB CYCLE & the ratio of instruction TLB miss page walk cycles to total cycles\\ \cline{2-4}
& 23 & DTLB MISS & misses in all levels of the data TLB per K instructions \\ \cline{2-4}
& 24 & DTLB CYCLE & data TLB miss page walk cycles to total cycles\\ \cline{2-4}
& 25 & DATA HIT STLB & DTLB first level misses that hit in the second level TLB per K instructions\\ \hline
Branch & 26 & BR MISS & branch miss prediction ratio \\ \cline{2-4}
Execution& 27 & BR EXE TO RE &  the ratio of executed branch instruction to retired branch execution\\ \hline
Pipeline Behavior & 28 & FETCH STALL & the ratio of instruction fetch stalled cycle to total cycles\\ \cline{2-4}
& 29 & ILD STALL & the ratio of Instruction Length Decoder stalled cycle to total cycles \\ \cline{2-4}
& 30 & DECODER STALL & the ratio of Decoder stalled cycles to total cycles \\ \cline{2-4}
& 31 & RAT STALL & the ratio of Register Allocation Table stalled cycles to total cycles\\ \cline{2-4}
& 32 & RESOURCE STALL & the ratio of resource related stalled to total cycles, which including load store buffer full stalls, Reservation Station full stalls, ReOrder buffer full stalls and etc \\ \cline{2-4}
& 33 & UOPS EXE CYCLE & the ratio of micro operation executed cycle to total cycles \\ \cline{2-4}
& 34 & UOPS STALL & the ratio of no micro operation executed cycle to total cycles \\ \hline
Offcore  & 35 & OFFCORE DATA & percentage of offcore data request\\ \cline{2-4}
Request& 36 & OFFCORE CODE & percentage of offcore code request\\ \cline{2-4}
& 37 & OFFCORE RFO & percentage of offcore Request For Ownership\\ \cline{2-4}
& 38 & OFFCORE WB& percentage of data write back to uncore\\ \hline
Snoop Response &39 & SNOOP HIT& HIT snoop responses per K instructions\\ \cline{2-4}
&40 & SNOOP HITE& HIT Exclusive snoop responses per K instructions\\ \cline{2-4}
&41 & SNOOP HITM& HIT Modified snoop responses per K instructions\\ \hline
Parallelism  & 42 & ILP & Instruction Level Parallelism\\ \cline{2-4}
& 43 & MLP & Memory Level Parallelism\\ \hline
Operation Intensity & 44 & INT TO MEM & integer computation to memory access ratio\\ \cline{2-4}
&45 & FP TO MEM& floating point computation to memory access ratio\\ \hline
\end{tabular}
\end{table}

\subsection{Removing Correlated Data}\label{static1}
Given the 32 workloads and 45 metrics for each workload, it is
difficult to analyze all the metrics to draw meaningful conclusions.
Note, however, that some metrics may be correlated.
For instance, long latency cache misses may cause pipeline stalls.
Correlated data can skew similarity analysis --- many
correlated metrics will overemphasize a particular property's importance.
So we eliminate correlated data before analysis.
Principle Component Analysis (PCA)~\cite{jolliffe2005principal} is
a common method for removing such correlated
data~\cite{phiansalkar2007analysisab,eeckhout2002workload,eeckhout2003quantifying,bienia2010fidelity}.
We first normalize metric values to a Gaussian distribution with
mean equal to zero and standard deviation equal to one
(to isolate the effects of the varying ranges of each dimension).
Then we use Kaiser's Criterion to choose the number of principle components (PCs).
That is, only the top few PCs, which have eigenvalues
greater than or equal to one, are kept. With Kaiser's Criterion,
the resulting data is ensured to be uncorrelated while capturing
most of the original information.

\subsection{Measuring Similarity} \label{static2}
We analyze the similarity among workloads
implemented with different software stacks.
Hierarchical clustering is one common way to perform such
analysis, for it can quantitatively show the similarity among workloads via a dendrogram.
Hierarchical clustering connects objects to form groups based on their distance.
In the beginning, each element is in a cluster of its own.
At each successive step, the two clusters
separated by the shortest distance are combined.
In the end all elements end up in the same cluster.

\subsection{Removing Redundancy} \label{k-means}
In order to generate a representative benchmark subset, we should
eliminate redundant workloads.
We use K-means clustering to group the workloads, and then we choose
a representative workload from each cluster.
We use the Bayesian Information Criterion (BIC) as a measure to evaluate the K-Means
efforts and choose the best K value.

\section{Experimental Setup} \label{exeevn}
We first describe the experimental environment for our study,
and then we explain how we obtain the microarchitectural-level data.

\subsection{Hardware Configurations} \label{hwcon}
We run all workloads on a five-node cluster.
A 1Gb ethernet network connects one master and four slaves.
Each node is equipped with two Intel Xeon E5645 (Westmere) processors and 32GB of memory.
These processors include six physical out-of-order cores with
speculative pipelines.
Table~\ref{hwconfigeration} lists the hardware details.
We disable hyperthreading and Turbo Boost because when enabled these features
makes it more complex to measure and interpret performance data~\cite{levinthal2009performance}.

\begin{table}
\caption{Details of the Hardware Configuration.}\label{hwconfigeration}
\scriptsize
\center
\begin{tabular}{|l|c|}
  \hline
  CPU Type & Intel \textregistered Xeon E5645\\ \hline
  \# Cores & 6 cores@2.4G \\ \hline
  \# Threads per Core& 1 thread \\ \hline
  \#Sockets & 2 \\ \hline
  \hline
  ITLB & 4-way set associative, 64 entries \\ \hline
  DTLB & 4-way set associative, 64 entries \\ \hline
  L2 Shared TLB& 4-way associative, 512 entries \\ \hline
  L1 DCache & 32KB, 8-way associative, 64 byte/line \\ \hline
  L1 ICache & 32KB, 4-way associative, 64 byte/line \\ \hline
  L2 Cache & 256 KB, 8-way associative, 64 byte/line \\ \hline
  L3 Cache &  12 MB, 16-way associative, 64 byte/line \\ \hline
  Memory & 32 GB , DDR3 \\  \hline
Hyper-Threading & Disabled \\ \hline
Turbo-Boost & Disabled \\ \hline
\end{tabular}
\end{table}

\subsection{Software Environments} \label{configuration}
We use the same system software configuration for both the Hadoop-based and Spark-based workloads.
Each cluster node runs Linux CentOS 6.4  with Linux kernel version 3.11.10.
The JDK version is 1.7.0.
The Hadoop and Spark versions are 1.0.2 and 0.8.1, respectively.
The Hive and Shark versions are 0.9.0 and 0.8.0, respectively.


\subsection{Performance Data Collection}
Most modern processors provide hardware performance monitoring counters (PMCs)
that track microarchitectural metrics.
In the Xeon processor, MSRs (Model Specific Registers)
can be set to specify which hardware events to count,
and an accompanying set of registers store those events' results.
We use Perf --- a profiling tool for Linux 2.6+ based systems~\cite{perf} ---
to specify the event numbers and corresponding unit masks for the MSRs.
We collect more than 50 events (some metrics require multiple events)
whose numbers and corresponding unit masks can
be found in the Intel Developer's Manual~\cite{guide2010intel}. Although Perf
can multiplex the PMCs, we run each workload multiple times to
obtain more accurate values for the metrics listed in Table~\ref{metric}.
We perform a ramp-up period for each application, and then collect performance data
throughout the lifetime of each workload.
We collect the data for all four slave nodes and take the mean.

\section{Software Stack Impact} \label{framework}
We employ PCA and hierarchical clustering to analyze the impact of
the software stack on the microarchitectural performance characteristics
exhibited by each workload.
We first give an overall similarity/dissimilarity analysis
based on hierarchical clustering.
We then project our workloads onto a principle component (PC) space to
investigate workload differences along different PC dimensions.
Finally, we demonstrate that a few microarchitectural-level metrics
suffice to effectively distinguish Hadoop-based and Spark-based workloads,
and we use these metrics to compare the differences.

\subsection {(Dis)similarity Analysis} \label{macro}

\begin{figure*}
\centering
\includegraphics[scale=0.6]{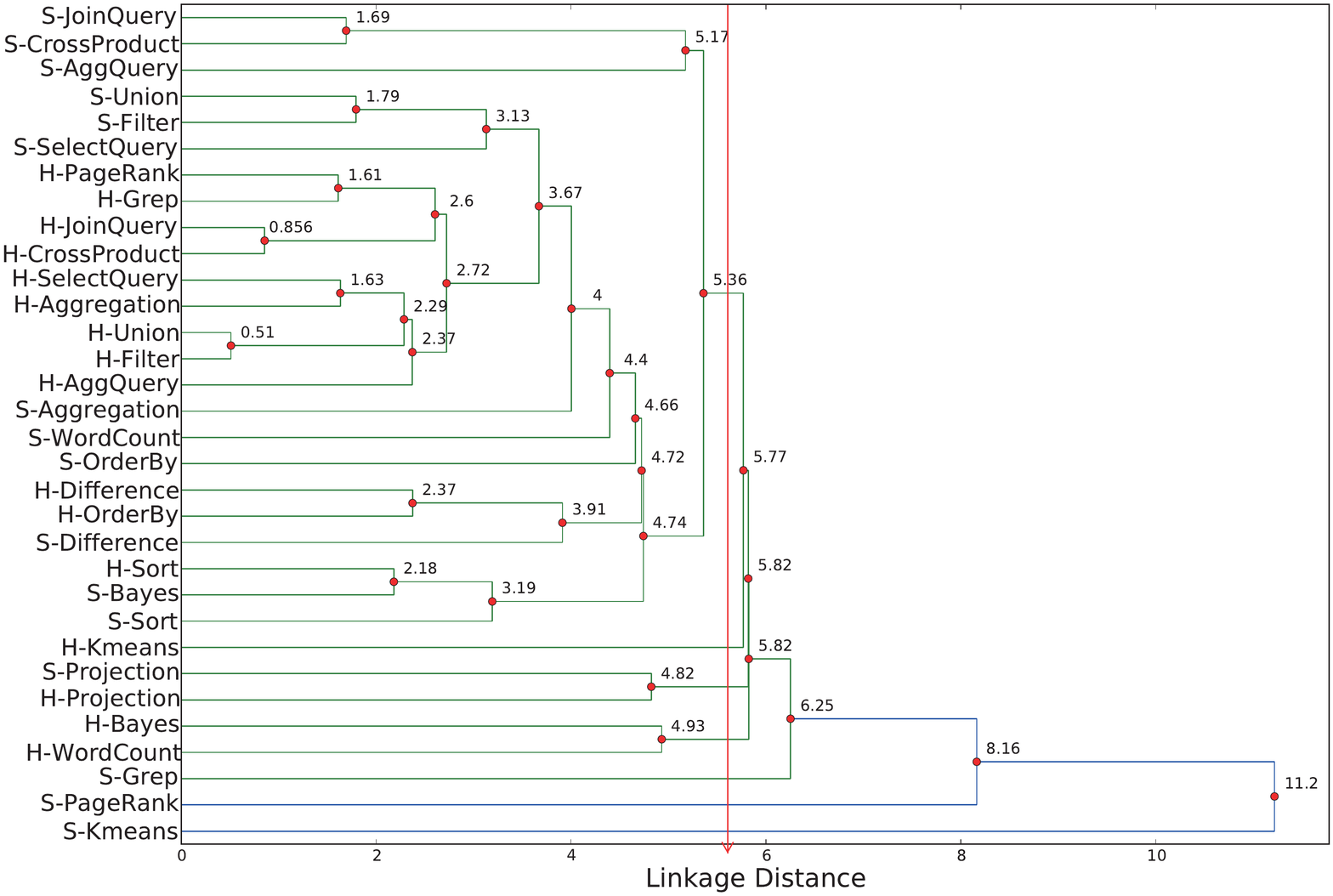}
\caption{Similarity of Hadoop (H) and Spark (S) workloads.}\label{Hclust-all}
\end{figure*}

Figure~\ref{Hclust-all} shows the dendrogram for all big data workloads
characterized in this paper. We create the dendrogram by applying PCA
and hierarchical clustering to the metrics in Table~\ref{metric}.
The dendrogram illustrates how each cluster is
composed by drawing a U-shaped link between a non-singleton cluster and its children.
The length of the top of the U-link is the distance between its
children. The shorter the distance, the more similar the children.
Like Phansalkar et al.~\cite{phiansalkar2007analysisab}, we use Euclidean distance.
Further, we use the single linkage distance to create the dendrogram.
That is to say the linkage distance between two clusters is made by a single
element pair, namely those two elements (one in each cluster) that are closest to each other.
In the figure, the x-axis shows the linkage distances among workloads.
In our study, we choose eight PCs that have eigenvalues greater
than one. These PCs retain 91.12\% variance.

We make the following observations from the figure.

\textbf{Observation 1:}
At the first clustering iteration, most (80\%) of clusters consist
of workloads that are based on the same software stack.
The only exceptions are S-Bayes/H-Sort and S-Projection/H-Projection
with linkage distances of 2.18 and 4.82, respectively. These linkage
distances exceed most of the distances between the children in the
clusters formed in the first iteration.
The shortest linkage distance between the workloads using different
software stacks to implement the same algorithm is 3.19 for H-Sort and S-Sort.

\textbf{Observation 2:}
Two workloads implementing the same algorithm on different software stacks
do not get clustered together in the first clustering iteration --- the only
exception is Projection. However, the linkage distance between H-Projection and S-Projection
is 4.82, which is large enough to indicate that their microarchitectural
behaviors are quite different.

\textbf{Observation 3:}
After the first iteration, workloads based on the same software stack are easier to cluster together,
e.g., S-Union, S-Filter, and S-SelectQuery workloads get clustered together
after just two iterations. Likewise, H-SelectQuery, H-Aggregation, H-Union, and H-Filter get
clustered after two iterations.

In the dendrogram, when workloads are clustered quickly with small linkage
distances, it indicates that the workloads are similar.
Workloads based on the same software stack are easier to cluster than programs that implement
the same algorithm but use different software stacks because the latter
have dramatically different behaviors.
We thus conclude that the software stacks have significant impacts on workload behaviors
--- even greater than that of the benchmark algorithms.
This phenomenon occurs because the implementations of the software stacks
allow programmers to develop applications without considering system
issues. For instance, a Hadoop-based WordCount application has only 50-some
lines of user application code in both the Map and Reduce functions.
In contrast, Hadoop's source code can be dozens or even hundreds of megabytes.
For example in Hadoop version 1.0.2, the size of the main source code (i.e. src folder size) is 67 MB.
The upshot is that the ratio of system software and middleware instructions
executed compared to user application instructions tends to be large,
which makes their impact on system behavior large, as well.

\textbf{Observation 4:}
H-Union/H-Filter, S-Union/S-Filter, H-JoinQuery/H-CrossProduct, and S-JoinQuery/S-CrossProduct,
are grouped, respectively, in the first clustering iteration.
On an algorithmic level, join query and cross product are similar.
Both operate records from two tables at once. Union and filter also perform similar
operations in the BigDataBench implementations. From this we conclude that
workloads using the same software stack to implement similar algorithms have similar behaviors.

\textbf{Observation 5:}
Most of the Hadoop-based workloads exhibit small linkage distances, and thus they are easier to cluster.
There are nine Hadoop workloads (H-Pagernak to H-AggQuery
in Figure~\ref{Hclust-all}) clustered within linkage distances of 2.72.
In contrast, only three Spark-based workloads (S-Union to
S-SelectQuery in Figure~\ref{Hclust-all}) are clustered within a similar distance
(3.13).
H-Union and H-Filter are clustered with
linkage distance of 0.51, and H-JoinQuery and H-CrossProduct are
clustered with distance of 0.856. In contrast, S-Union and
S-Filter get grouped with linkage distance of 1.79, and S-JoinQuery
and S-CrossProduct are clustered with distance of 1.69.

The above observation shows that Hadoop-based workloads have
more similarities with each other than their Spark-based counterparts.
We thus conclude that Hadoop has a larger impact on microarchitectural behaviors
than Spark. Its software stack dominates application behavior, minimizing the
impact of potentially diverse behaviors introduced by user application code.
Spark is simpler than Hadoop with respect to number of lines of code, and thus
it dominates system behavior less. For instance, in contrast to
Hadoop's source code size, Spark's whole folder is only 11 MB for version 0.81.

The bottom line is that software stacks play an important role in big data systems,
and that impact may increase over time.
As more and more mechanisms and services are added,
software stacks become more and more complicated. Taking the tar package
as an example, its size in Hadoop version 1.0.2 was 32.6MB,
whereas in version 2.4, its size grows to 133MB.
The impact of software stacks to application behaviors may become greater and greater.
Thus to choose representative workloads to benchmark big data systems,
the software stack should be considered separately from the algorithms in user application code,
and different software stacks should be included for thorough workload
characterization.

\subsection{Principle Component Space Analysis} \label{pcspace}
\begin{figure}
\centering
\includegraphics[scale=0.28]{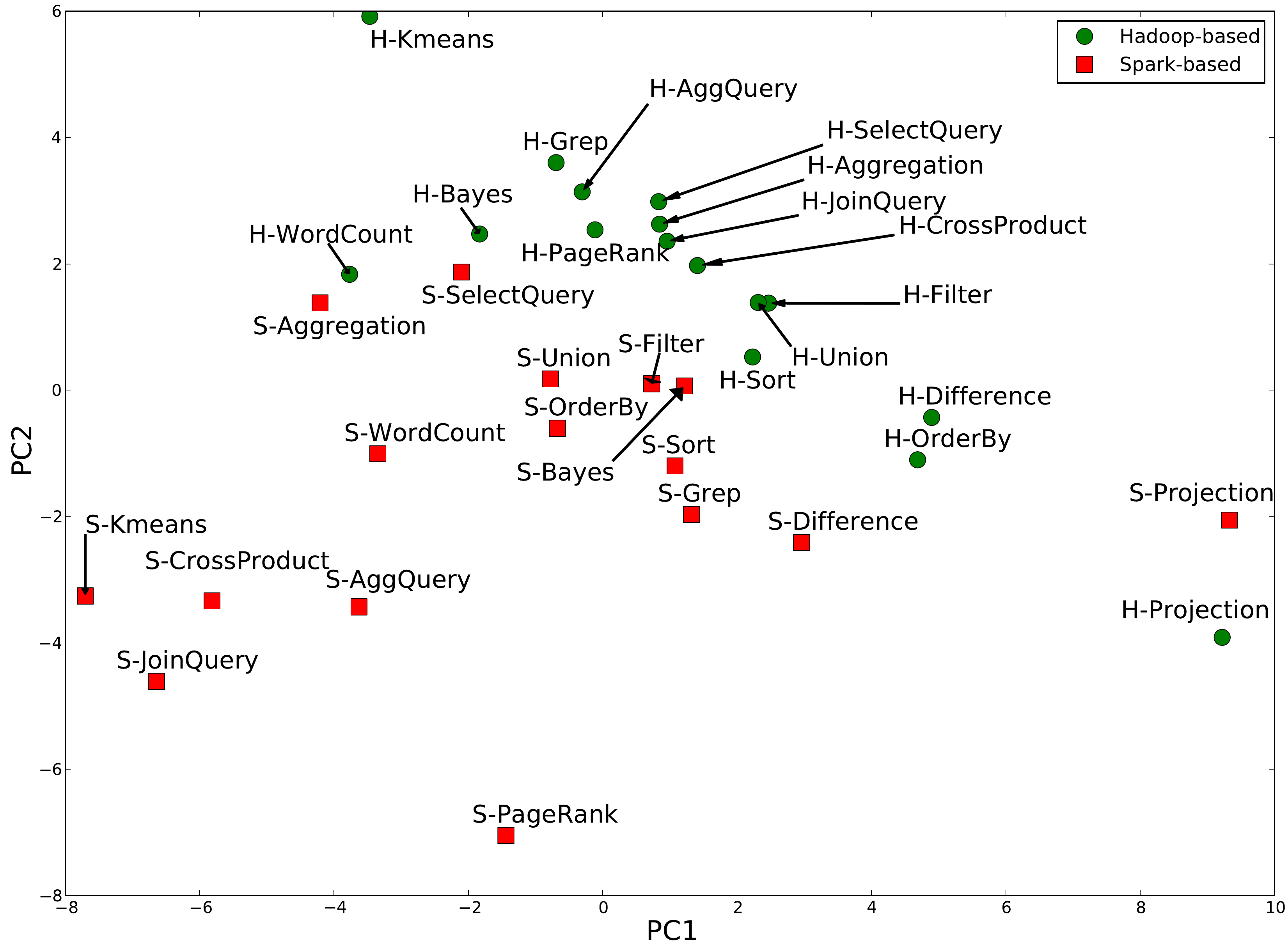}
\caption{Scatter plot of workloads using the first and second principle components}\label{PC-all1}
\end{figure}

\begin{figure}
\centering
\includegraphics[scale=0.28]{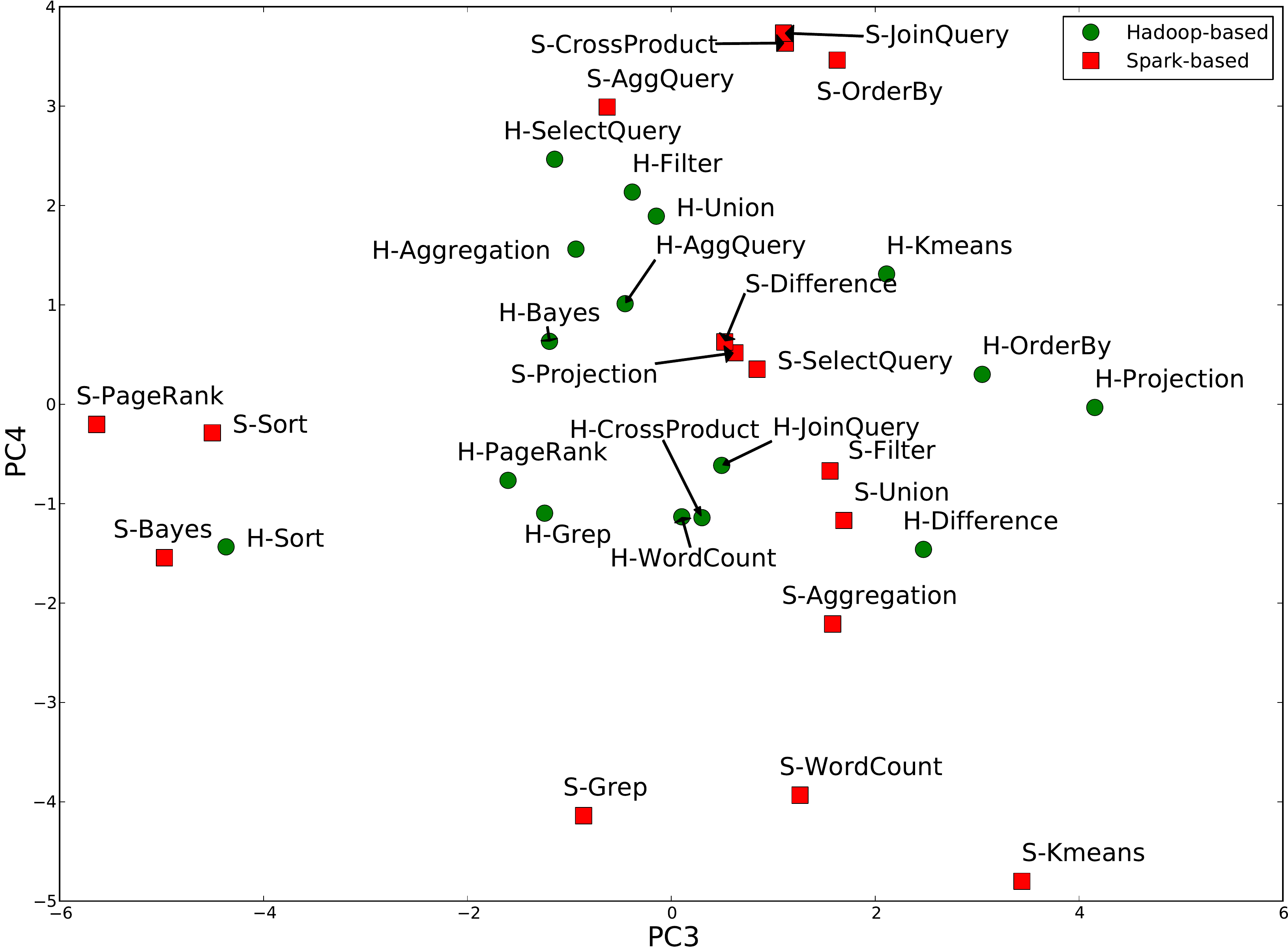}
\caption{Scatter plot of workloads using the third and fourth principle components}\label{PC-all2}
\end{figure}

In this section, we project the workloads into principle component spaces.
We only present the first four PCs covering 71.45\% of
the total variance due to space limitations.
Figure~\ref{PC-all1} shows the scatter plot of the first and second principle components,
PC1 and PC2. Figure~\ref{PC-all2} shows this for PC3 and PC4.
These visualizations only include a subset of available information
based on the first four PCs. Workloads appearing close in these figures
may in fact be farther away when all PCs are considered.
Even in this subset, however, workloads that appear far apart indeed
exhibit significantly different behaviors.

\begin{figure*}
\centering
\includegraphics[scale=0.48]{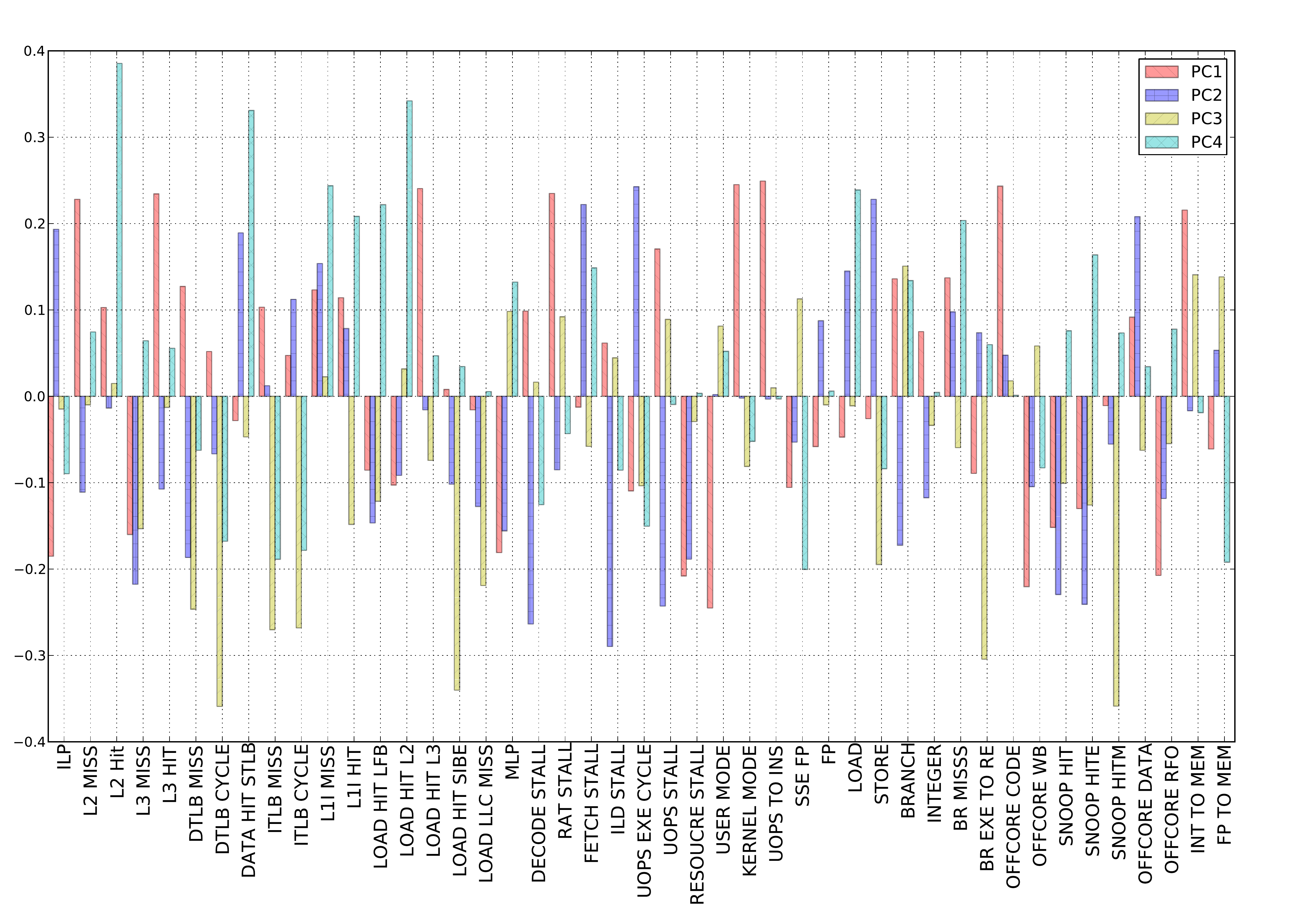}
\caption{Factor loadings for all workloads}\label{factor}
\end{figure*}

The factor loadings can show the map from the original microarchitecture level metrics to the principle components.
Figure~\ref{factor} shows the factor loadings of the first four PCs.
This means that the PC1 is computed
as $PC1 = -0.18 \times ILP + 0.23 \times L2~MISS+ 0.102 \times L2~HIT + ...$.
PC1 is positively dominated by L2 MISS,
L3 HIT, LOAD HIT L3, RAT STALL, KERNEL MODE, UOPS TO INS, OFFCORE CODE, and INT TO MEM.
And PC1 is also negatively dominated by RESOURCE STALL, USER MODE, OFFCORE WB, and OFFCORE RFO.
That is to say,
when compared with other workloads, the workloads in Figure~\ref{PC-all1} with a high value along PC1, have more L2 MISS,
L3 HIT, LOAD HIT L3, RAT STALL, KERNEL MODE instructions, OFFCORE CODE requests, higher UOPS TO INS ratio and INT TO MEM ratio. At the same time they have less RESOURCE STALL, USER MODE instructions, OFFCORE WB, and OFFCORE RFO requests.
Similarly, Figure~\ref{factor} also shows the factors
that positively or negatively dominate the other three PCs.
The factor loadings can also help us to understand the behavior differences among workloads.
Such as H-Kmeans has highest value along PC2 (Figure~\ref{PC-all1}) for it has more STORE instructions, FETCH STALL, and DATA HIT STLB.
S-Kmeans, S-WordCount, and S-Grep discriminate themselves along PC4 (Figure~\ref{PC-all2}) due to their low L2 HIT rates, STLB HIT
rates, L1 instruction hit rates, percentages of LOAD instructions,
and branch miss rates and their high ITLB miss rates
and more ITLB walk cycles. Factor loadings also explain the other
isolated points in a similar way, but we omit them for brevity.

The Spark-based workloads are spread widely along the x-axis (PC1) in Figure~\ref{PC-all1}.
In contrast, the Hadoop-based workloads are grouped in the middle of the chart.
The Spark-based workloads thus exhibit more diversity than their Hadoop-based counterparts
with respect to PC1.
Figure~\ref{PC-all2} displays similar phenomena. The Spark-based workloads
nearly cover all of the space for PC3 and PC4, whereas the Hadoop-based workloads
are again grouped in the center.
There are some isolated points in the figures, most of which represent
Spark-based workloads exhibiting behaviors that differ
from the other workloads.
These phenomena verify our finding in Section~\ref{macro} that the
Spark-based workloads show more diversity than the Hadoop-based workloads
with respect to microarchitectural behaviors
because the Hadoop software stack minimizes the impact of the user application code.


\subsection{Differentiating Hadoop and Spark} \label{dis}
On the y-axis (PC2) in Figure~\ref{PC-all1} we find that
most of the Hadoop-based workloads are located towards the top,
whereas most of Spark-based workloads are located in the middle or bottom
parts of the chart.  Spark-based workloads and Hadoop-based workloads are mixed along other PC dimensions.
So we conclude that PC2 is the main component that differentiates
the Hadoop-based workloads from the Spark-based workloads.
We therefore investigate which microarchitectural-level metrics
dominate PC2 values.
Figure~\ref{factor} shows the weights of each microarchitecture metric for PC1-PC4.
The metrics that negatively dominate PC2's values are L3 MISS,
DTLB MISS, DECODE STALL, ILD STALL, UOPS STALL, RESOURCE STALL, BRANCH, SNOOP HIT,
and SNOOP HITE.
The metrics that positively dominate its values are ILP, DAT HIT STLB,
FETCH STALL, UOPS EXE CYCLE, STORE, and OFFCORE DATA.
We therefore compare the workloads using those metrics.

\begin{figure}
\centering
\includegraphics[scale=0.67]{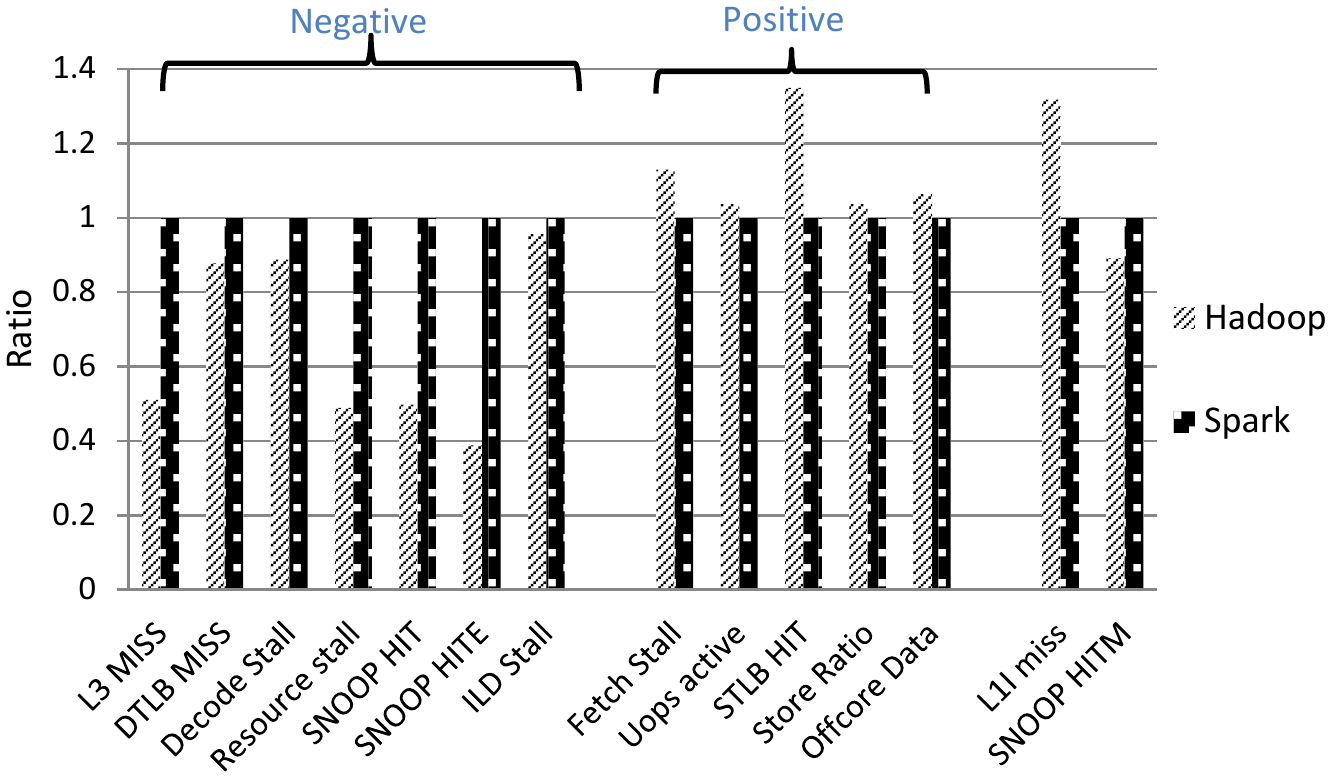}
\caption{Metrics causing Hadoop and Spark to behave differently}\label{PC2data}
\end{figure}

Figure~\ref{PC2data} shows the main metrics that dominate PC2's values.
We calculate the mean for each metric for each kind of big workload
and present normalized Hadoop-based values using the Spark-based workloads as the baseline.
In the figure, the metrics that have negative impact have lower values
for the Hadoop-based workloads, whereas those with positive impact have
higher values. Most of the Hadoop-based workloads have higher PC2 values than
their Spark-based counterparts.

We make the
following observations from Figure~\ref{PC2data}.

\textbf{Observation 6:}
The Spark-based workloads have a large amount of L3 cache misses per kilo instructions,
about twice those of the Hadoop-based workloads.

\textbf{Observation 7:}
The Hadoop-based workloads have more data shared TLB hits and
fewer DTLB misses than the Spark-based workloads.

\textbf{Observation 8:}
The Hadoop-based workloads have more instruction fetch stalls,
whereas the Spark-based workloads have more resource related stalls.

We can divide the processor architecture into two parts:
the in-order frontend, which fetches, decodes, and issues instructions, and
the out-of-order backend, which executes instructions and writes data back
to the register file.
The instruction fetch stalls belong to the frontend and the resource
related stalls belong to the backend.
The former are mainly caused by L1 instruction cache misses.
Figure~\ref{PC2data} shows that the L1 I-cache misses per kilo instructions (MPKI)
are about 30\% higher for the Hadoop-based workloads on average.

Most resource stalls (e.g., load/store buffer full and reservation station full)
are likely caused by long latency L3 data cache or data TLB misses.
Note that in our processor, an STLB (second level Shared TLB) hit
implies an L1 TLB miss.
In Figure~\ref{PC2data}, the DTLB miss metric includes both levels.
The Hadoop-based workloads have more data STLB hits, which means
that the data STLB is more efficient for those workloads.
We confirm that by calculating the data STLB hit rates.
The data STLB hit rate is 61.48\%, on average, for Hadoop-based workloads,
whereas the rate for their Spark-based counterparts is 50.80\%, on average.

In general, the large number of L3 cache and data TLB misses cause more backend stalls for
the Spark-based workloads, and the high instruction cache misses
cause more frontend stalls for the Hadoop-based workloads.
We conclude that the Hadoop-based workloads have larger instruction footprints
than their Spark-based counterparts.
These phenomena could be because larger application binaries
(including the software stacks) create more I-cache
misses~\cite{jiacharacterization,ferdman2011clearing}, especially
in the case of Hadoop.
However the Spark-based workloads have larger data footprints than Hadoop-based workloads.
This may be caused by the different operation modes on intermediate data between Hadoop and Spark software stacks.

\textbf{Observation 9:}
The Spark-based workloads have more SNOOP HIT and SNOOP HITE
responses than Hadoop-based workloads.

Both responses represent coherence traffic among different cores, so they
reflect data sharing patterns.
The Spark-based workloads also have more SNOOP HITM
responses. 
These metrics show that Spark-based workloads have more data
sharing among different cores.

As the number of cores increases, the traffic caused by the cache
coherence protocol will increase.
Programmers and architects should thus
pay much attention to data sharing (particularly false sharing)
in Spark-based workloads.
This implies that different software stacks will require different
optimization strategies or lead to different system design decisions:
there is no single solution.

\section{Subsetting}
Computer architects often use benchmark suite subsets
that capture the most important system behaviors~\cite{yi2006evaluating}.
This approach reduces simulation and evaluation time, and thus shortens the research period.
A well selected subset can reduce workload redundancy while keeping representativity.
Here we use the data obtained from Section~\ref{framework} to
eliminate redundant workloads and generate a representative subset
of BigDataBench.
Note that both the previous section's similarity analyses and this
section's subsetting are all from a computer architecture point of
view. Results may differ if subsetting is performed
from a different point of view~\cite{yoo2007hierarchical}.

Statistical approaches are frequently used to facilitate
workload subsetting. In this section we use PCA and clustering analysis to remove the
redundant workloads similarly to previous
efforts~\cite{eeckhout2003quantifying,phiansalkar2007analysisab,hoste2006comparing}.

\subsection{Clustering}
We use clustering on the eight principle components obtained from
the PCA algorithm to group workloads into similarly behaving application
clusters (the eight PCs are those in Section~\ref{framework}).
In particular, we use K-Means clustering for a number of K values.
Inspired by previous research~\cite{hoste2006comparing,sherwood2002automatically},
we use the Bayesian Information Criterion (BIC) to choose the proper K value.
The BIC is a measure of the ``goodness of fit'' of a clustering for a
data set. The larger the BIC scores, the higher the probability that
the clustering is a good fit to the data.
Here we determine the \emph{K} value that yields the highest BIC score.
We use the formulation from Pelleg et al.~\cite{pelleg2000x} shown in Equation~\ref{equ1}
to calculate the BIC.

\begin{equation} \label{equ1}
BIC(D,K)=l(D|K)-\frac{{p}_{j}}{2}log(R)
\end{equation}

Where $D$ is the original data set to be clustered.
In this paper, $D$ is $32 \times 8$ matrix which indicates $32$ workloads and each workload is represented by $8$ PCs (Principle Components).
$l(D|K)$ is the likelihood.
$R$ is the number of workloads to be clustered.
$p_j$ is the sum of $K-1$ class probabilities, which is $K+dK$.
$d$ is the dimension of each workloads, which is $8$ in this paper for we choose $8$ PCs.
To compute $l(D|K)$, we use Equation~\ref{equ2}.

\begin{equation} \label{equ2}
\begin{aligned}
l(D|K)=\sum{^{K}_{i=1}}(-\frac{{R}_{i}}{2}log(2\pi) -\frac{{R}_{i}\cdot d}{2}log(\sigma^2) \\
-\frac{R_i-K}{2}+R_ilogR_i-R_ilogR)
\end{aligned}
\end{equation}

Where $R_i$ is the number of points in the $i^{th}$ cluster, and $\sigma^2$ is
the average variance of the Euclidean distance from each point to its cluster center,
which is calculate by Equation~\ref{equ3}.

\begin{equation} \label{equ3}
\sigma^2=\frac{1}{R-K}\sum_i(x_i-\mu (i))^2
\end{equation}

Here $x_i$ is the data point assigned to cluster $i$, and $\mu(i)$ represents
the center coordinates of cluster $i$.

We ultimately cluster the 32 workloads into seven groups which are listed in Table~\ref{kresult}. Please note that though the result of K-Means is similar to that of hierarchical clustering in Figure~\ref{Hclust-all},  the K-Means clustering can not measure  the similarity among different workloads quantitatively like the hierarchical clustering.
For example, the hierarchical clustering measures that  H-Projection and S-Projection are clustered together with a linkage distance of 4.82, while the K-Means can only indicate that they are clustered  together qualitatively.

\begin{table}
\caption{The result of K-Means Clustering Algorithm}\label{kresult}
\centering
\begin{tabular}{|c|p{6cm}|c|}
\hline
Cluster & Workloads & Number \\ \hline
1 & H-PageRank, H-Grep, H-JoinQuery, H-Sort, H-CrossProduct, S-Bayes, S-Grep, S-Sort &8 \\ \hline
2& H-AggQuery, S-Filter, S-Union, S-SelectQuery, S-OrderBy, H-Kmeans & 6\\\hline
3& S-Aggregation, S-WordCount, S-Kmeans, H-Wordcount, H-Bayes &5 \\\hline
4& H-OrderBy, H-Difference, S-Difference, S-PageRank & 4 \\\hline
5& H-Aggregation, H-Filter, H-Union, H-SelectQuery &4 \\ \hline
6& S-CrossProduct, S-JoinQuery, S-AggQuery & 3 \\ \hline
7& H-Projection, S-Projection & 2\\ \hline
\end{tabular}
\end{table}

\subsection{Selecting Representative Workloads}
The representative for each cluster can be chosen by two approaches,
as mentioned by Eeckhout et al.~\cite{eeckhout2003quantifying}.
The first is to choose the workload that is as close as possible to
the center of the cluster it belongs to. The other is to select an
extreme workload situated at the ``boundary'' of each cluster.
We experiment with both approaches.
Table~\ref{represent} lists the representative workloads
selected from each cluster by both approaches.

The third column of Table~\ref{represent} gives the maximal linkage distance among
representative workloads for each approach.
We find that workloads chosen by the first approach
do not cover workloads with long linkage distances, e.g., S-PageRank.
The maximal distance among representative workloads selected
by the first approach is smaller than the maximal distance among
those selected by the second approach.
This means that workloads chosen by the first method are not as diverse as the second approach does.

In Section~\ref{framework} we perform  hierarchical clustering to analyze workload
similarity.
To use hierarchical clustering to select seven representative workloads,
we just draw a vertical line at a point close to 5.6 for the vertical line is intersecting with seven horizontal lines in Figure \ref{Hclust-all}.
There are four single workload consisted clusters have intersections with the vertical line.
This reveals that we should choose those four workloads,
namely, S-PageRank, S-Kmeans, S-Grep, and H-Kmeans.
These four workloads are also chosen by the second approach,
whereas none of them is chosen by the first approach.
We therefore deem the second approach superior in selecting representative workloads.
And the rationale behind this method can be that the behaviors of other workloads in the cluster can be extracted from the behavior of the boundary one, e.g. through interpolation ~\cite{eeckhout2003quantifying}.
\begin{table}
\caption{Representative workloads chosen by different approaches}\label{represent}
\center
\begin{tabular}{|c|l|c|}
  \hline
  Approach & Representatives Workloads \footnotemark[1] & Maximal Linkage Distance \\ \hline
  Nearest to  &H-PageRank (8)  & \\ \cline{2-2}
   Cluster Center              & S-Select (6)&  \\ \cline{2-2}
                 &S-Aggregation (5)& \\ \cline{2-2}
                 &S-Difference (4)& 5.82\\ \cline{2-2}
                 &H-Aggregation (4)& \\ \cline{2-2}
                 &S-CrossProduct (3)& \\ \cline{2-2}
                 &S-Projection (2)& \\ \hline
  Farthest from  & S-Grep (8)& \\ \cline{2-2}
   Cluster Center & H-Kmeans (6) & \\ \cline{2-2}
                  &S-Kmeans (5)& \\ \cline{2-2}
                  &S-PageRank (4)& 11.20\\ \cline{2-2}
                  &H-SelectQuery (4) &\\ \cline{2-2}
                  &S-AggQuery (3)&\\ \cline{2-2}
                  &H-Projection (2)&\\
  \hline
\end{tabular}
\footnotemark[1]{The number of workloads that the selected workloads can represent are given in parentheses.}
\end{table}

\begin{figure}
\centering
\includegraphics[scale=0.45]{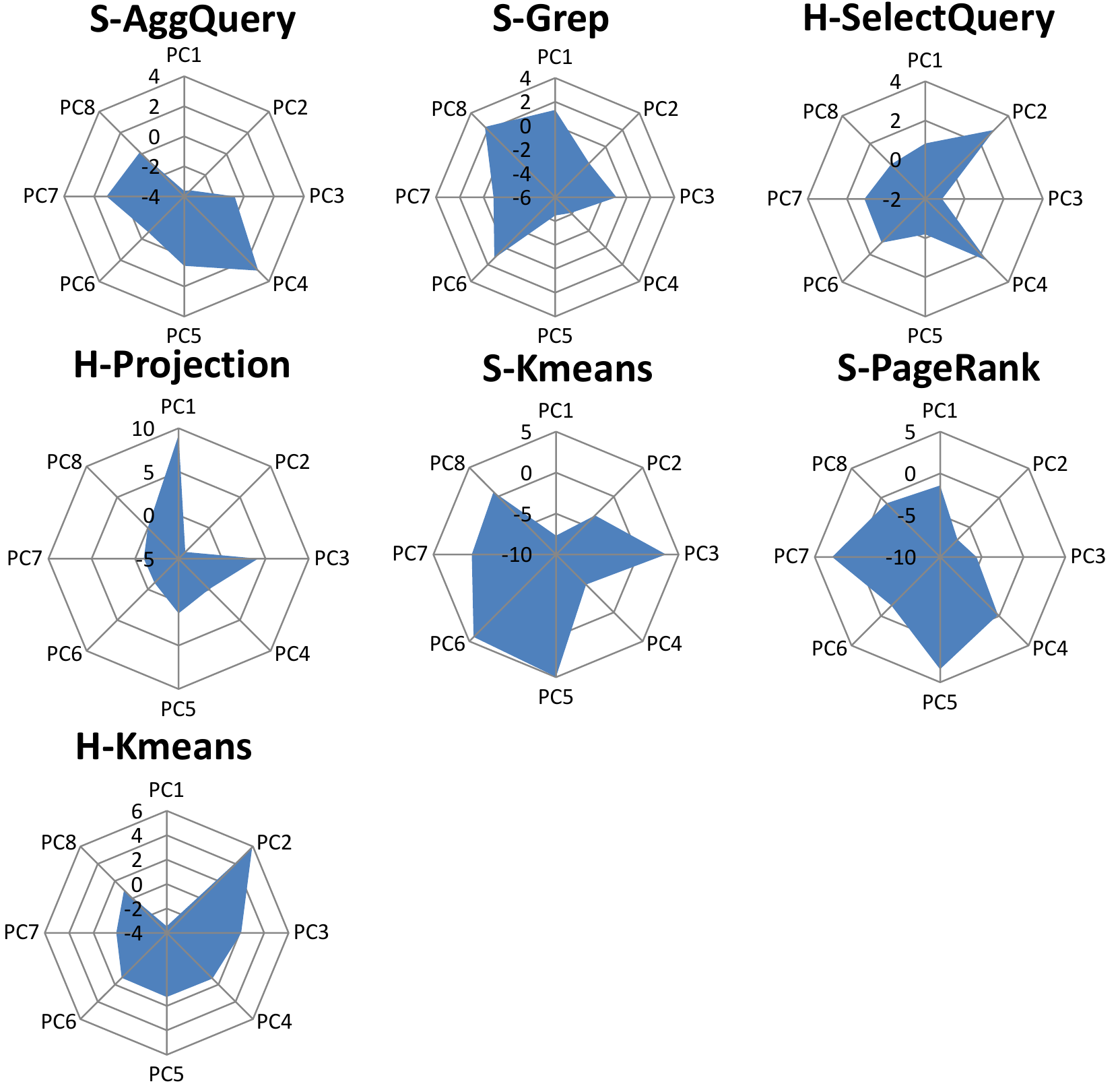}
\caption{Kiviat diagrams of the representative workloads}\label{kiviat}
\end{figure}

Figure~\ref{kiviat} shows Kiviat diagrams for the representative workloads chosen by the second approach.
The diagrams illustrate that the representative workloads
are diverse and that different workloads are dominated by
different principle components.

\subsection{BigDataBench Simulator Version}
We deploy the representative applications on a full-system simulator and
release the simulator images as the BigDataBench simulator version.
More information can be found on the project web page \url{http://prof.ict.ac.cn/BigDataBench/simulatorversion/}.
For the big data workloads change frequently~\cite{barroso2009datacenter}, the state-of-art workloads and software stacks will be integrated into and out-of-date ones will be removed from BigDataBench.
We will continue the subsetting work, so the BigDataBench simulator version on the web site may be different from the one mentioned in this paper when you access the web site.
It is our hope that this simulator version will benefit architecture researchers
by reducing the costs of evaluating alternative technological approaches
in big data systems.
\section{Related Work}

Wang et al.~\cite{wang2014bigdatabench} propose a comprehensive big data workloads suite, including different structured, un-structured, and semi-structured data. However, a large number of benchmarks pose great challenges, since our usual simulation-based research
methods become prohibitively expensive.
Xi et al.~\cite{xi2011characterization} characterize micro architecture level behaviors of the search engine.
Ferdman et al.~\cite{ferdman2011clearing} select seven scale-out workloads according to popularity. They use hardware performance
counters to analyze microarchitectural behaviors of those scale-out workloads.
They compare the scale-out workloads and traditional benchmarks to identify
the key contributors to the microarchitecture inefficiency on modern processors.
They conclude that mismatches exist between the needs of scale-out workloads
and the capabilities of modern processors.
Jia et al.~\cite{jiacharacterization,jia2014implications} characterize microarchitectural
characteristics of data analysis workloads, also finding that they exhibit
different characteristics from traditional workloads.
Luo et al.~\cite{luo2012cloudrank} compare two clusters' performance and power consumption using hybrid big data workloads. Continuing the work in ~\cite{luo2012cloudrank}, our group also releases the multi-tenancy version of BigDataBench, which support the scenarios of multiple tenants running heterogeneous applications in the same datacenter.  The multi-tenancy version of BigDataBench is publicly available from~\cite{mutil-tenancy}, which is helpful for the research of datacenter resource management and other interesting issues~\cite{wang2012cloud,zhan2013cost}.


Much work focuses on comparing the performance of
different data management systems. For OLTP or database systems evaluation,
TPC-C~\cite{tpcc} is often used to evaluate transaction-processing system
performance in terms of transactions per minute. Cooper et al.~\cite{cooper2010benchmarking}
define a core set of benchmarks and report throughput and latency
results for five widely used data management systems. Pavlo et al.~\cite{pavlo2009comparison}
compare the MapReduce paradigm to traditional parallel DBMS platforms.
The Berkeley AMPLab Big Data Benchmark~\cite{AMP} can also be used to compare
MapReduce and parallel DBMS platforms in terms of response time for
a handful of relational queries across different data sizes.

Many prior studies characterize benchmarks via statistical data analysis techniques
such as PCA and clustering. For instance, Eeckhout et
al.~\cite{eeckhout2003quantifying,eeckhout2002workload},
Phansalkar et al.~\cite{phiansalkar2007analysisab,phansalkar2007subsetting}, and Yi et al.~\cite{yi2006evaluating}
all employ statistical analyses to characterize workloads, analyze redundancy,
and perform subsetting for general-purpose and embedded benchmark suites.
Our work here is inspired by their methods, although we focus on a very
different application domain.

\section{Conclusion}
Achieving benchmark comprehensiveness while enabling efficient experimentation
raises great challenges for the architecture community, in general, and the
problem is particularly acute in big data systems research.
The diversity of big data software stacks aggravates the challenge,
especially for simulation-based research.
Here we use 45 microarchitectural metrics to characterize 32 big data workloads
on two different software stacks.
Our statistical analyses on the gathered metrics reveal that software stacks
have significant impacts on workload behavior, and we find that in the context
of our experiments, this impact is greater than that of the algorithms employed
in user application code.

In order to facilitate simulation-based research, we create a
streamlined, representative subset of the benchmarks in the BigDataBench
suite. To do so, we remove the redundant workloads via K-Means clustering.
In order to choose the best K value for the K-Means algorithm, we
use BIC as the statistical test method to evaluate the K-Means
efforts.
After subsetting, we are left with some representative workloads.
Our hope is that our subsetting approach and resulting benchmark
suite will enable more architecture researchers to study
alternative organizations and technologies for big data systems.

\section*{Acknowledgment}
We are very grateful to anonymous reviewers.
This work is supported
in part by the State Key Development
Program for Basic Research of China (Grant No.2011CB302502 and 2014CB340402) and the NSFC project (Grant No.61202075).

\bibliographystyle{IEEEtran}
\bibliography{IEEEabrv,tex}
\end{document}